\documentclass[twocolumn,floatfix,aps,prb,eqsecnum]{revtex4-2}
\usepackage{graphicx}
\usepackage{amsmath}
\usepackage{amssymb}
\usepackage{bm}
\usepackage{hyperref}

\begin{document}

\title{Phase-shifted Andreev levels in an altermagnet Josephson junction}
\author{C. W. J. Beenakker}
\affiliation{Instituut-Lorentz, Universiteit Leiden, P.O. Box 9506, 2300 RA Leiden, The Netherlands}
\author{T. Vakhtel}
\affiliation{Instituut-Lorentz, Universiteit Leiden, P.O. Box 9506, 2300 RA Leiden, The Netherlands}
\date{June 2023}

\begin{abstract}
We compute the effect of a \textit{d}-wave magnetization (altermagnetism) on the spectrum of bound states (Andreev levels) in a junction between two \textit{s}-wave superconductors (gap $\Delta_0$, phase difference $\phi$). Compared to a nonmagnetic junction, the $\phi$-dependence of the spectrum is shifted by an offset $\pm\delta\phi$, dependent on the spin direction, so that the Andreev levels become spin-polarized. In a planar junction, oriented along the crystalline axis of $d_{xy}$-wave symmetry, the excitation energies are determined by the normal-state transmission probability $T$ according to $E=\Delta_0\sqrt{1-T\sin^2\tfrac{1}{2}(\phi\pm\delta\phi)}$. We calculate the corresponding Josephson energy and supercurrent, recovering the 0--$\pi$ transition of related studies.
\end{abstract}
\maketitle

\section{Introduction}

Altermagnets (metals with a \textit{d}-wave magnetization that ``alternates'' direction in momentum space) differ from ferromagnets and antiferromagnets in that they combine a spin-polarized Fermi surface with a vanishing net magnetization \cite{Sme22a, Sme22b,Maz22a, Maz23}. Candidate altermagnetic materials include $\text{RuO}_2$, MnTe, and $\text{Mn}_5\text{Si}_3$ \cite{Fen22,Occ22,Gon23,Sme22c}. The interplay of altermagnetism and superconductivity produces unusual effects \cite{Maz22b}, including orientation-dependent Andreev reflection \cite{Sun23,Pap23}, negative critical supercurrent with finite-momentum Cooper pairing \cite{Oua23,Zha23}, and topological Majorana modes \cite{Zhu23,Gho23}.

\begin{figure}[tb]
\centerline{\includegraphics[width=0.8\linewidth]{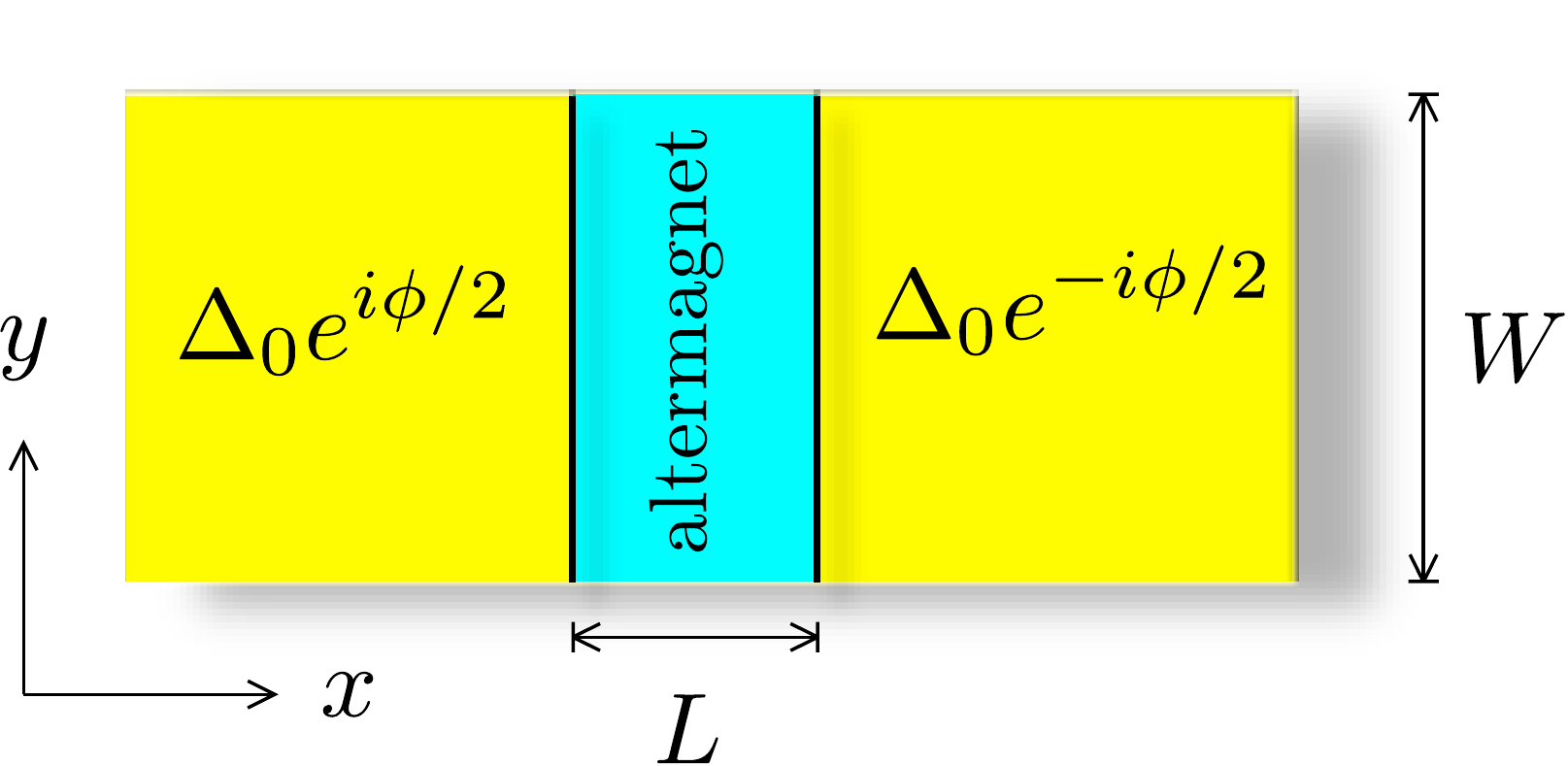}}
\caption{Josephson junction consisting of a pair of superconductors (pair potential $\Delta_0$, phase difference $\phi$) connected by an altermagnet. We consider a short planar junction, $W\gg \xi_0\gg L$.
}
\label{fig_layout}
\end{figure}

A basic building block for these effects is the altermagnet Josephson junction, in which two \textit{s}-wave superconductors (gap $\Delta_0$, phase difference $\phi$) are connected by a \textit{d}-wave magnetic region (see Fig.\ \ref{fig_layout}). The subgap excitations are Andreev levels, electron-hole superpositions confined to the junction. If the length $L$ of the junction is short compared to the superconducting coherence length $\xi_0=\hbar v_{\rm F}/\Delta_0$, there is one Andreev level per spin direction and per transverse mode.

For a non-magnetic Josephson junction the Andreev levels are spin-degenerate, and the $\phi$-dependence of the excitation energy is given by \cite{Bee91}
\begin{equation}
E=\Delta_0\sqrt{1-T\sin^2(\phi/2)},\label{EphiT}
\end{equation}
where $T\in(0,1)$ is the transmission probability through the junction of an electron mode at the Fermi level in the normal state.

Here we investigate how the altermagnet modifies the excitation spectrum. For unit transmission through a planar junction the $\cos(\phi/2)$ Andreev band is split into spin-polarized branches. The splitting is a phase shift $\pm\delta\phi$ that depends on the angle $\theta$ of the junction with the crystalline axis of $d_{xy}$-wave symmetry. For $\theta=0$ the relation \eqref{EphiT} with $\phi\mapsto \phi\pm\delta\phi$ still holds for non-unit transmission. We test these analytical predictions with a computer simulation of a tight-binding model of the altermagnet Josephson junction.

These results provide an alternative description of the 0--$\pi$ transition reported recently \cite{Oua23,Zha23}, where the sign of the critical current oscillates with increasing $L$. The description is particularly simple for $\theta=0$, when the phase shift $\delta\phi$ is proportional to the transverse momentum $k_y$, so $\partial E/\partial\phi\propto \partial E/\partial k_y$. The supercurrent $I\propto \int_{-k_{\rm F}}^{k_{\rm F}} dk_y (dE/d\phi)$ is therefore directly given by an energy difference --- the integral operation cancels the derivative. The resulting critical current $I_c$ oscillates $\propto (\sin \delta\phi_{\rm max})/\delta\phi_{\rm max}$, with $\delta\phi_{\rm max}\propto k_{\rm F}L$.

\section{Altermagnet Josephson junction}

We consider the Josephson junction geometry of Fig.\ \ref{fig_layout}, consisting of a pair of superconducting regions ($x<0$ and $x>L$) connected by a non-superconducting magnetic metal ($0<x<L$). The superconducting pair potential $\Delta$ has \textit{s}-wave symmetry, while the magnetization has the \textit{d}-wave symmetry characteristic of an altermagnet. The junction has length $L$ and width $W$. 

For large width $W$, and without impurity scattering, we may assume translational invariance in the $y$-direction, so that the transverse momentum $k_y$ is a good quantum number. We work in the short-junction regime $L\ll\xi_0$.

The excitation spectrum is described by the Bogoliubov-De Gennes (BdG) Hamiltonian
\begin{align}
{\cal H}(\bm{k})={}&\begin{pmatrix}
H_0(\bm{k})&\Delta\\
\Delta^\ast&-\sigma_yH_0^\ast(-\bm{k})\sigma_y
\end{pmatrix},\label{HBdGdef}\\
H_0(\bm{k})={}&\frac{\hbar^2}{2m}(k_x^2+k_y^2)-\mu\nonumber\\
&+\frac{\hbar^2}{m}t_{1}k_xk_y\sigma_z+\frac{\hbar^2}{m}t_{2}(k_y^2-k_x^2)\sigma_z.\label{H0def}
\end{align}
The $\sigma_\alpha$'s are Pauli spin matrices, $\bm{k}=(k_x,k_y)$ is the electron momentum (two-dimensional, for simplicity), and $\mu=\hbar^2 k_{\rm F}^2/2m=\tfrac{1}{2}mv_{\rm F}^2$ is the Fermi energy. In what follows we set $\hbar=1$ and the electron mass $m=1$ (restoring units in the final results).

The $d$-wave exchange interaction is characterised by two dimensionless parameters $t_1$ and $t_2$, which depend on the angle $\theta$ of the altermagnet-superconductor (AS) interface relative to the crystalline axes,
\begin{equation}
t_{1}=2t_{0}\cos 2\theta,\;\;t_{2}=t_0\sin 2\theta.
\end{equation}
The parameter $t_0$ is of order $10^{-1}$ \cite{Sme22c}. For $\theta=0$ the magnetization has pure $d_{xy}$-wave symmetry, for $\theta=\pi/4$ it has pure $d_{x^2-y^2}$-wave symmetry.

The $4\times 4$ Hamiltonian \eqref{HBdGdef} decouples into $2\times 2$ blocks ${\cal H}_\uparrow$ and ${\cal H}_\downarrow$. The blocks are spin-polarized in the sense that electrons and holes occupy opposite spin bands, so that each block describes quasiparticles with a definite magnetic moment.

We consider the spin-up block 
\begin{align}
&{\cal H}_\uparrow(\bm{k})=\begin{pmatrix}
H_+(\bm{k})&\Delta\\
\Delta^\ast&-H_-(\bm{k})
\end{pmatrix},\\
&H_\pm(\bm{k})=\tfrac{1}{2}(k_x^2+k_y^2)-\mu\pm t_1 k_xk_y\pm t_2 (k_y^2-k_x^2).
\end{align}
The spin-down block ${\cal H}_\downarrow$ is obtained by switching $t_1\mapsto-t_1$, $t_2\mapsto -t_2$. 

Near the Fermi level ($E=0$) we may linearize the $k_x$-dependence of ${\cal H}_\uparrow$ at given momentum $k_y$, parallel to the AS interfaces at $x=0$ and $x=L$. In the altermagnet region $0<x<L$, where $\Delta=0$, we have
\begin{equation}
{\cal H}_\uparrow=(\bar{v}-\tau_z\delta v)\nu_z\tau_z(k_x-Q_0- Q_z\tau_z).\label{H+linear}
\end{equation}
The Pauli matrix $\tau_z$ acts on the electron-hole degree of freedom, while $\nu_z$ distinguishes right-movers from left-movers. 

In Eq.\ \eqref{H+linear} we have introduced the velocities
\begin{equation}
v_\pm=v_{\rm F}\sqrt{1\pm 2t_2-(k_y/k_{\rm F})^2(1-t_1^2-4t_2^2)}\equiv\bar{v}\pm\delta v,\label{vbardef}
\end{equation}
and momentum offsets
\begin{equation}
\begin{split}
&Q_0=k_{\rm F}(1-4t_2^2)^{-1}\bigl(\nu_z(\bar{v}-2t_2\delta v)/v_{\rm F}-2t_1t_2k_y/k_{\rm F}\bigr),\\
&Q_z=k_{\rm F}(1-4t_2^2)^{-1}\bigr(\nu_z(2t_2\bar{v}-\delta v)/v_{\rm F}-t_1k_y/k_{\rm F}\bigr).
\end{split}
\end{equation}
For later use we also define
\begin{equation}
\begin{split}
&Q_0^\pm=k_{\rm F}(1-4t_2^2)^{-1}\bigl(\pm(\bar{v}-2t_2\delta v)/v_{\rm F}-2t_1t_2k_y/k_{\rm F}\bigr),\\
&Q_z^\pm=k_{\rm F}(1-4t_2^2)^{-1}\bigr(\pm(2t_2\bar{v}-\delta v)/v_{\rm F}-t_1k_y/k_{\rm F}\bigr).
\end{split}\label{Q0zpm}
\end{equation}

\section{Andreev levels without normal reflection}

Andreev reflection at $x=0$ and $x=L$ converts electrons into holes, with absorption of the missing charge of $2e$ into the superconducting condensate \cite{And64}. It coexists with normal reflection, without charge conversion. In this section we neglect normal reflections, we will include these in the next section. 

Andreev reflection from electron to hole with energy $E$, at a pair potential $\Delta=\Delta_0e^{i\phi}$, is associated with a phase shift $e^{-i\phi-i\alpha(E)}$, where
\begin{equation}
\alpha(E)=\arccos(E/\Delta_0)\in(0,\pi),\;\;|E|<\Delta_0.\label{alphadef}
\end{equation}
The phase shift for reflection from hole to electron is $e^{+i\phi-i\alpha(E)}$. We set $\Delta=\Delta_0 e^{i\phi/2}$ at the left superconductor ($x<0$) and $\Delta=\Delta_0e^{-i\phi/2}$ at the right superconductor ($x>L$). 

The condition for a bound state is that the phase increment on a round-trip $x=0\mapsto L\mapsto 0$ is a multiple of $2\pi$. For $k_{\rm F}L\ll \mu/\Delta_0$ (equivalently, $L\ll\xi_0$, the short-junction regime) we may ignore the energy dependence of the phase shift accumulated in the normal region, while retaining the energy dependence of the Andreev reflection phase shift $\alpha(E)$. This gives the bound-state condition
\begin{equation}
\phi+2LQ_z^\pm=\pm 2\alpha(E)\;\text{mod}\;(2\pi).
\end{equation}
The $\pm$ sign distinguishes whether the right-moving quasiparticle is an electron or a hole. The contribution $Q_0$ to the phase shift cancels in the round-trip, only the increment $Q_z$ contributes.

We thus obtain two branches of Andreev levels $E^\pm_\uparrow$, with
\begin{subequations}
\label{Euparrow}
\begin{align}
E_\uparrow^\pm={}&\pm\Delta_0\,\text{sign}\,(\sin\psi_\uparrow^\pm)|\cos\tfrac{1}{2}\psi_\uparrow^\pm|,\\
&\psi_\uparrow^\pm=\phi+2LQ_z^\pm.
\end{align}
\end{subequations}
We have added the subscript $\uparrow$ as a reminder that these are the bound states of ${\cal H}_\uparrow$. For ${\cal H}_\downarrow$ one replaces $L\mapsto-L$.

\begin{figure}[tb]
\centerline{\includegraphics[width=0.8\linewidth]{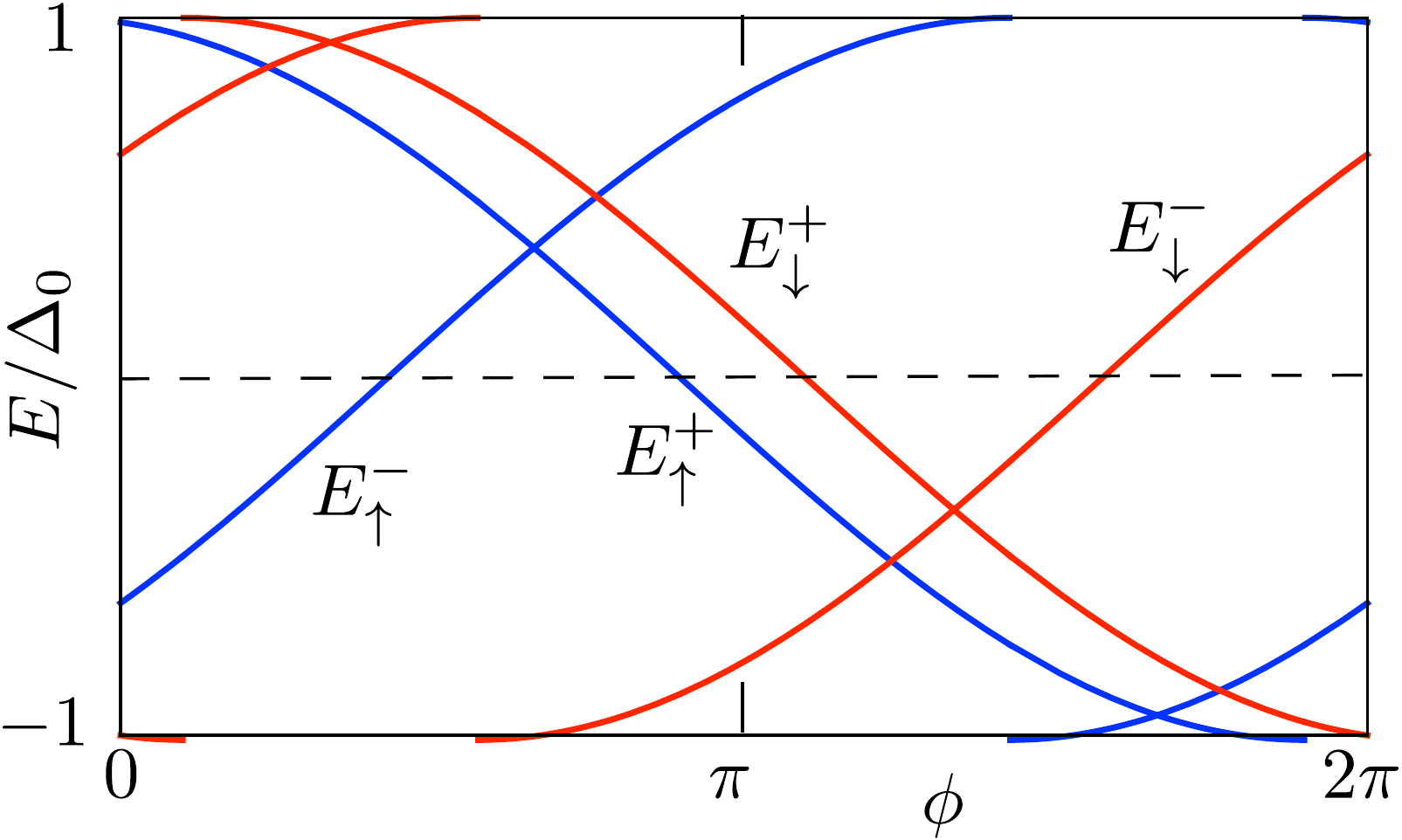}}
\caption{Phase dependence of the Andreev levels, computed from Eq.\ \eqref{Euparrow} for $t_1=t_2=0.1$, $k_{\rm F}L=20$, $k_y/k_{\rm F}=1/2$. There are four levels at each value of the phase difference, distinguished by the spin direction $\uparrow,\downarrow$ and by the $\pm$ sign of the current $\propto dE/d\phi$ which they carry. Another four levels at $k_y/k_{\rm F}=-1/2$ ensures the electron-hole symmetry of the spectrum. Normal reflections are neglected in this calculation.
}
\label{fig_levels}
\end{figure}

In a non-magnetic Josephson junction the Andreev levels are spin degenerate, with a cosine phase dependence \cite{Kul77}: $E=\pm \Delta_0\cos(\phi/2)$. As illustrated in Fig.\ \ref{fig_levels}, the altermagnet breaks up the cosine into branches that are phase shifted by a spin-dependent amount. Each branch connects the edges of the gap at $\pm\Delta_0$, where the bound states merge with the continuous spectrum. Electron-hole symmetry ($\pm E$ symmetry of the spectrum) is ensured by  the identity
\begin{equation}
Q_z^+(k_y)=-Q_z^-(-k_y)\Rightarrow E_\uparrow^\pm(k_y)=-E_\downarrow^\mp(-k_y).\label{Qpmsymmetry}
\end{equation}

\section{Including normal reflection}

An electron incident on the superconductor may be Andreev reflected as a hole, but it may also be reflected as an electron. Such normal reflection can be modeled by the insertion of a tunnel barrier at the two ends of the altermagnet. We assume that the barrier potential $V(x)$ does not break the translational invariance along the $y$-direction, so that the transverse momentum $k_y$ remains a good quantum number. We also assume the potential is spin-independent.

A simple solution of the scattering problem is possible for pure $d_{xy}$-wave pairing ($t_2=0$). The BdG Hamiltonian then reads
\begin{align}
{\cal H}={}&\bigl[\tfrac{1}{2}k_x^2+\tfrac{1}{2}k_y^2+V(x)-\mu\bigr]\tau_z\nonumber\\
&+\tfrac{1}{2}\bigl[t_{1}(x)k_x+k_xt_1(x)\bigr]k_y\sigma_z\nonumber\\
&+\Delta_0(x)\bigl[\tau_x\cos\phi(x)-\tau_y\sin\phi(x)\bigr].
\end{align}
The $x$-dependence of the magnetization and pair potential is included to describe the entire junction profile.

We make the unitary transformation ${\cal H}\mapsto U(x){\cal H}U^\dagger(x)$, with
\begin{equation}
U(x)=\exp\left(i\tau_z\sigma_zk_y\int_0^x t_1(x')dx'\right),
\end{equation}
resulting in
\begin{align}
{\cal H}={}&\bigl(\tfrac{1}{2}k_x^2+\tfrac{1}{2}(1-t_1^2)k_y^2+V(x)-\mu\bigr)\tau_z\nonumber\\
&+\Delta_0(x)\bigl(\tau_x\cos\tilde{\phi}(x)-\tau_y\sin\tilde{\phi}(x)\bigr),\label{Hunitary}
\end{align}
where $\tilde\phi(x)=\phi(x)+2\sigma_z k_y\int_0^x t_1(x')\,dx'$ is a spin-dependent phase difference.

So for a given $k_y$ and given spin direction, the altermagnet Josephson junction is equivalent to a nonmagnetic Josephson junction with phase difference $\phi\pm 2k_yL t_1$. The factor $1-t_1^2$ that multiplies $k_y^2$ in Eq.\ \eqref{Hunitary} amounts to an anisotropic mass, this factor can be set to unity for $t_1\ll 1$. We can then use the known result \eqref{EphiT} for the Andreev levels in a nonmagnetic Josephson junction,
\begin{equation}
\label{Ebarrier}
\begin{split}
&E_\uparrow^\pm(k_y)=\pm\Delta_0\sqrt{1-T(k_y) \sin^2(\phi/2- k_yL t_1)},\\
&E_\downarrow^\pm(k_y)=\pm\Delta_0\sqrt{1-T(k_y) \sin^2(\phi/2+ k_yL t_1)},
\end{split}
\end{equation}
where $T(k_y)$ is the transmission probability through the junction in the normal state ($\Delta_0=0$).

If $t_2$ is nonzero we do not have such a closed-form and general expression for the Andreev levels. We specify to the case of a tunnel barrier at each AS interface, with tunnel probability $\Gamma$ (the same at $x=0$ and at $x=L$). The scattering matrix calculation in App.\ \eqref{app_barrier} gives the spin-up Andreev levels $E_\uparrow^\pm$ as the two solutions of the nonlinear equation
\begin{widetext}
\begin{align}
(1-\Gamma)^2 \cos\bigl(2 \alpha(E)+L (Q_z^+-Q_z^-)\bigr)+\cos\bigl(2 \alpha(E)-L (Q_z^+-Q_z^-)\bigr)-\Gamma^2 \cos\bigl(\phi+L (Q_z^++Q_z^-)\bigr)\nonumber\\
=2 (1-\Gamma) \cos\bigl(L (Q_z^+-Q_z^-)\bigr)+4 (1-\Gamma)  \cos\bigl(L (Q_0^+-Q_0^-)\bigr)(1-E^2/\Delta_0^2).\label{Andreevbarrier}
\end{align}
\end{widetext}
The spin-down Andreev levels $E_\downarrow^\pm$ are the solutions of Eq.\ \eqref{Andreevbarrier} upon replacement of $L$ by $-L$.

As a check, if we now set $t_2=0$ we have $Q_z^+=Q_z^-$, $Q_0^+=-Q_0^-$, and Eq.\ \eqref{Andreevbarrier} has the solution \eqref{Ebarrier} with 
\begin{equation}
T(k_y)=\frac{\Gamma^2}{2 (1-\Gamma) \cos 2 L Q_0^++1+(1-\Gamma)^2}.\label{Tbarrier}
\end{equation}
This is indeed the normal-state transmission probability through a double-barrier junction, at momentum $Q_0^+=\sqrt{k_{\rm F}^2-(1-t_1^2)k_y^2}$.

\begin{figure}[tb]
\centerline{\includegraphics[width=0.8\linewidth]{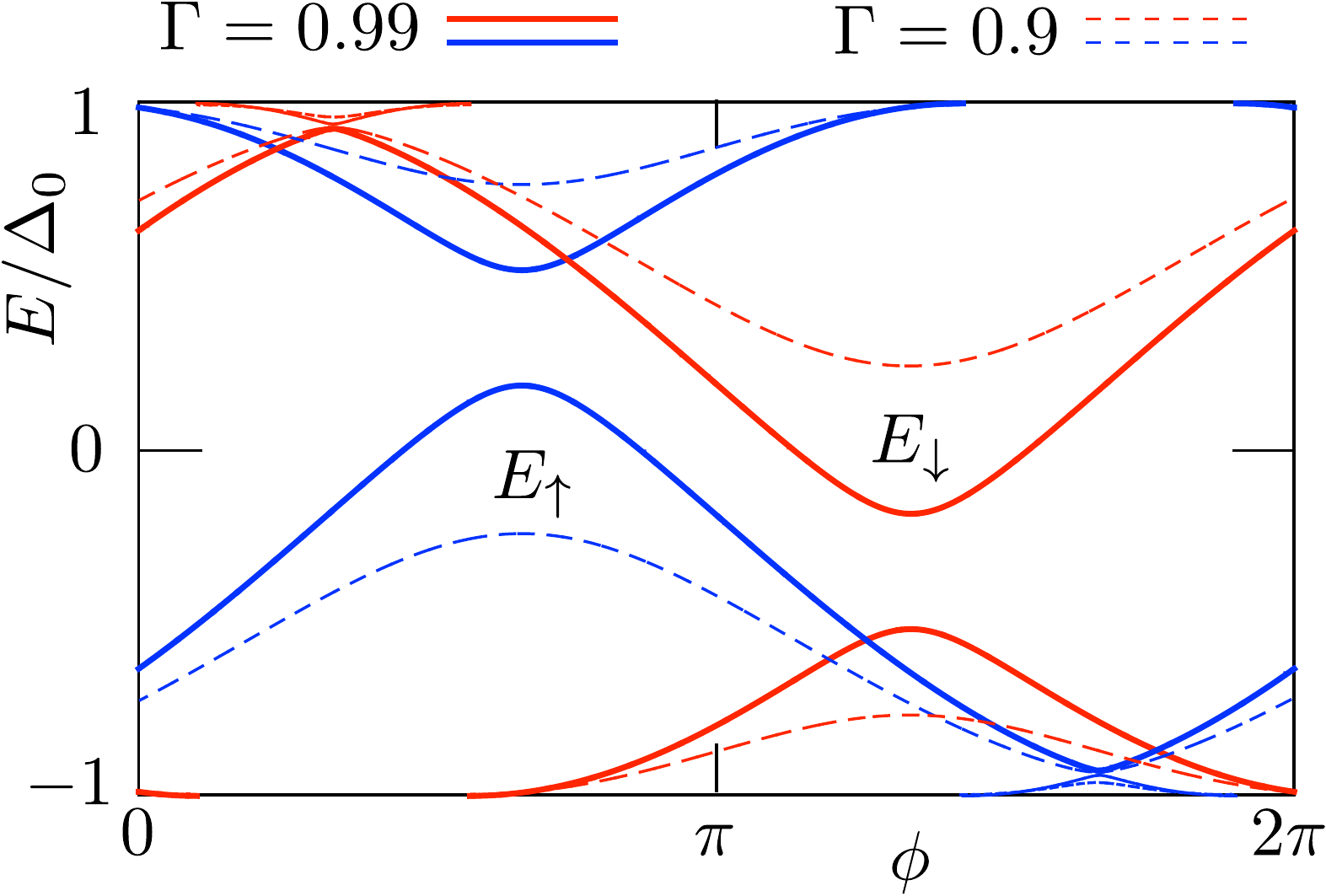}}
\caption{Same as Fig.\ \ref{fig_levels}, but now including normal reflections at each AS interface. The spectra are calculated from Eq.\ \eqref{Andreevbarrier} for two values of the transmission probability $\Gamma$, at $t_1=t_2=0.1$, $k_{\rm F}L=20$, $k_y/k_{\rm F}=1/2$.
}
\label{fig_levelsbarrier}
\end{figure}

If $t_2\neq 0$ Eq.\ \eqref{Andreevbarrier} can readily be solved numerically. As illustrated in Fig.\ \ref{fig_levelsbarrier}, crossings in the spectrum between levels of the same spin become anticrossings.

\section{Comparison with computer simulations}

To test these analytical predictions we have discretized the BdG Hamiltonian \eqref{H0def} on a square lattice and computed the subgap excitation spectrum numerically (see App. \ref{app_simulations}). These computer simulations fully include the normal reflections at the AS interfaces and they do not rely on the short-junction approximation. 

\begin{figure}[tb]
\centerline{\includegraphics[width=0.6\linewidth]{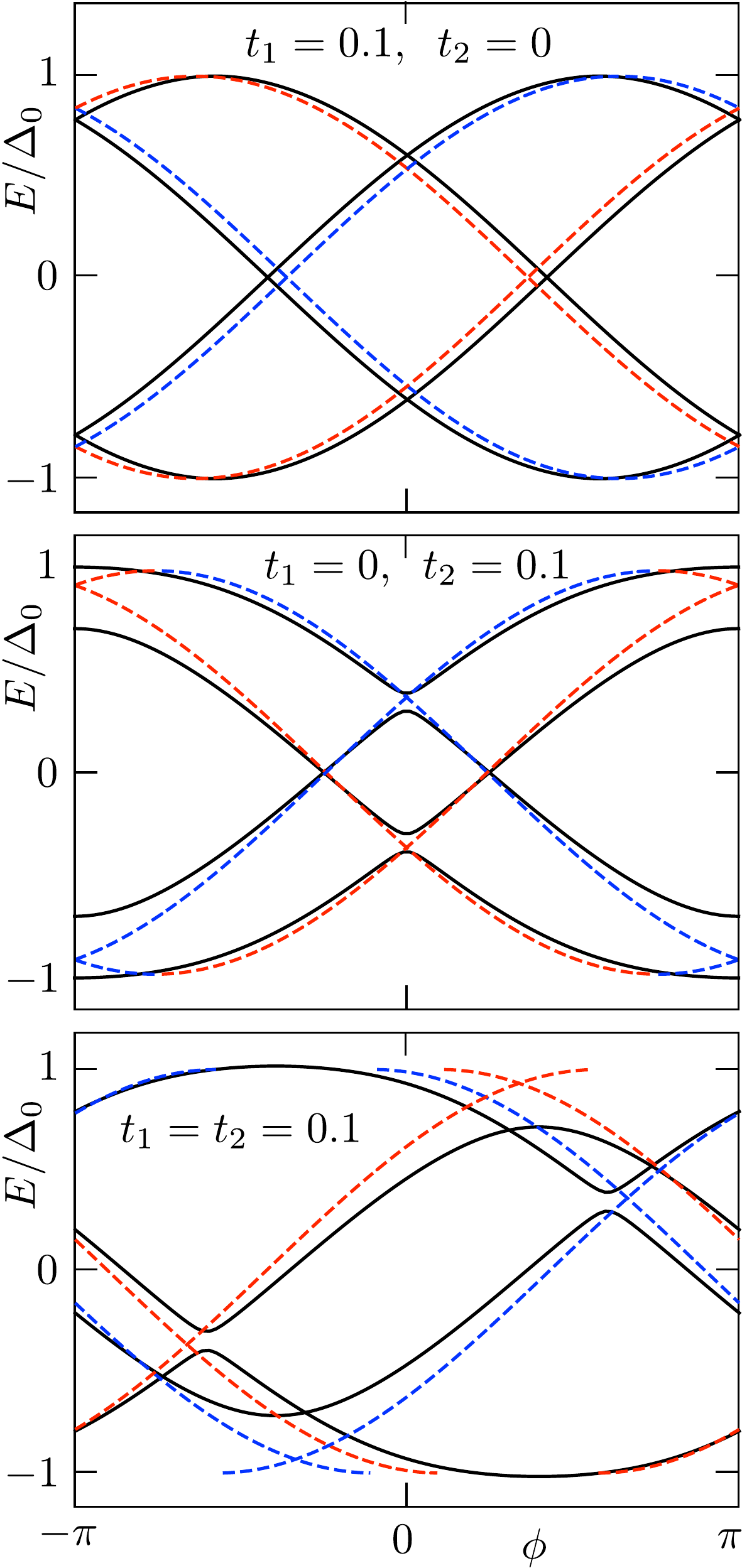}}
\caption{Andreev level spectra in the altermagnet Josephson junction, for $k_{\rm F}L=20$, $k_y/k_{\rm F}=1/2$, and different choices of $t_1,t_2$. The solid curves result from the numerical solution of the BdG equation on a lattice. The dashed curves are the analytical predictions \eqref{Euparrow}, in which normal reflections at the AS interfaces are neglected (blue for spin-up, red for spin-down).
}
\label{fig_simulations}
\end{figure}

In Fig.\ \ref{fig_simulations} we compare with the analytical predictions that ignore normal reflection. As expected, the main effect of normal reflection at the AS interfaces is to transform the crossings between same-spin branches into anticrossings. The effect is most pronounced when $t_1$ and $t_2$ are both nonzero: In the two cases of pure $d_{xy}$-wave or pure $d_{x^2-y^2}$-wave magnetization the crossings are only weakly affected.

\begin{figure}[tb]
\centerline{\includegraphics[width=0.8\linewidth]{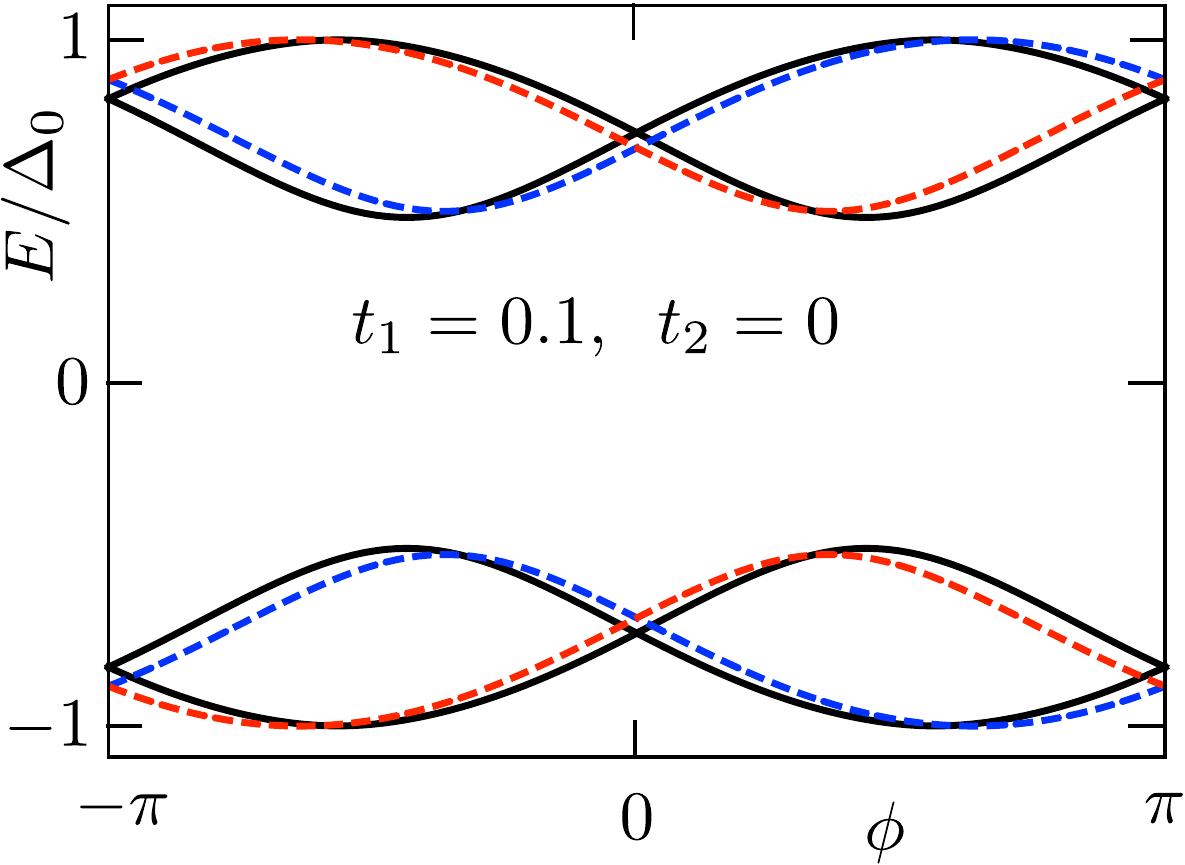}}
\caption{Solid curves: Andreev levels for $d_{xy}$-wave pairing in the presence of a tunnel barrier at the two AS interfaces ($k_{\rm F}L=20$, $k_y/k_{\rm F}=1/2$). The normal-state transmission probability $T=0.75$ through the junction was obtained directly from the computer simulation (by setting $\Delta_0\equiv 0$). The dashed curves are the analytical prediction \eqref{Ebarrier}, for the same value of $T$ (blue for spin-up, red for spin-down).
}
\label{fig_test}
\end{figure}

In Fig.\ \ref{fig_test} we test the relation \eqref{Ebarrier} between the Andreev levels and the normal-state transmission probability in the case of $d_{xy}$-wave pairing. The agreement is quite good, without any adjustable parameter.

\section{Josephson energy and supercurrent}

The supercurrent in the short-junction regime is carried entirely by the bound states, the continuous spectrum does not contribute \cite{Bee91,Bee92}. In equilibrium at inverse temperature $\beta$ the supercurrent $I$ is given by the phase derivative of the Josephson energy $F$,
\begin{equation}
I=\frac{2e}{\hbar}\frac{d}{d\phi}F,\;\;
F=-\sum_{E>0}\tfrac{1}{2}E\tanh\bigl(\tfrac{1}{2}\beta E\bigr),
\end{equation}
where $\sum_{E>0}$ is a sum over the transverse momentum $k_y$ and spin $\uparrow,\downarrow$ of the Andreev levels in the interval $(0,\Delta_0)$. 

In the absence of normal reflections, we find from Eqs.\ \eqref{Euparrow} and \eqref{Qpmsymmetry} that
\begin{subequations}
\label{Fdef}
\begin{align}
F={}&-\sum_{k_y}\sum_{s=\pm} \tfrac{1}{2}\varepsilon_s\Delta_0\tanh(\tfrac{1}{2}\varepsilon_s\beta \Delta_0),\\
&\varepsilon_s=\bigl|\cos(\phi/2+s LQ_z^+)\bigr|.
\end{align}
\end{subequations}
The transverse momenta range over the interval $(-k_{\rm max},k_{\rm max})$, with 
\begin{equation}
k_{\rm max}=k_{\rm F}\sqrt{\frac{1- 2t_2}{1-t_1^2-4t_2^2}}.
\end{equation}
in view of Eq.\ \eqref{vbardef}. In a junction of width $W\gg L,\xi_0$, and at zero temperature, one has
\begin{equation}
I(\phi)=-\frac{e\Delta_0}{\hbar}\frac{W}{2\pi}\int_{-k_{\text{max}}}^{k_{\text{max}}}dk_y\,\frac{d}{d\phi}(\varepsilon_++\varepsilon_-).\label{Iphiintegral}
\end{equation}

The integral over $k_y$ in Eq.\ \eqref{Iphiintegral} can be carried out in closed form for the case $t_2=0$ of a pure $d_{xy}$-wave magnetization, when $\varepsilon_s=|\cos(\phi/2-st_1k_yL)|$. We find
\begin{equation}
I=\frac{I_0}{2\alpha}\biggl(|\cos(\phi/2-\alpha)|-|\cos(\phi/2+\alpha)|\biggr),\label{Inoreflection}
\end{equation}
with $I_0=(e\Delta_0/\hbar)(k_{\text{max}}W/\pi)$ and $\alpha=t_1 k_{\text{max}}L$. 

\begin{figure}[tb]
\centerline{\includegraphics[width=0.8\linewidth]{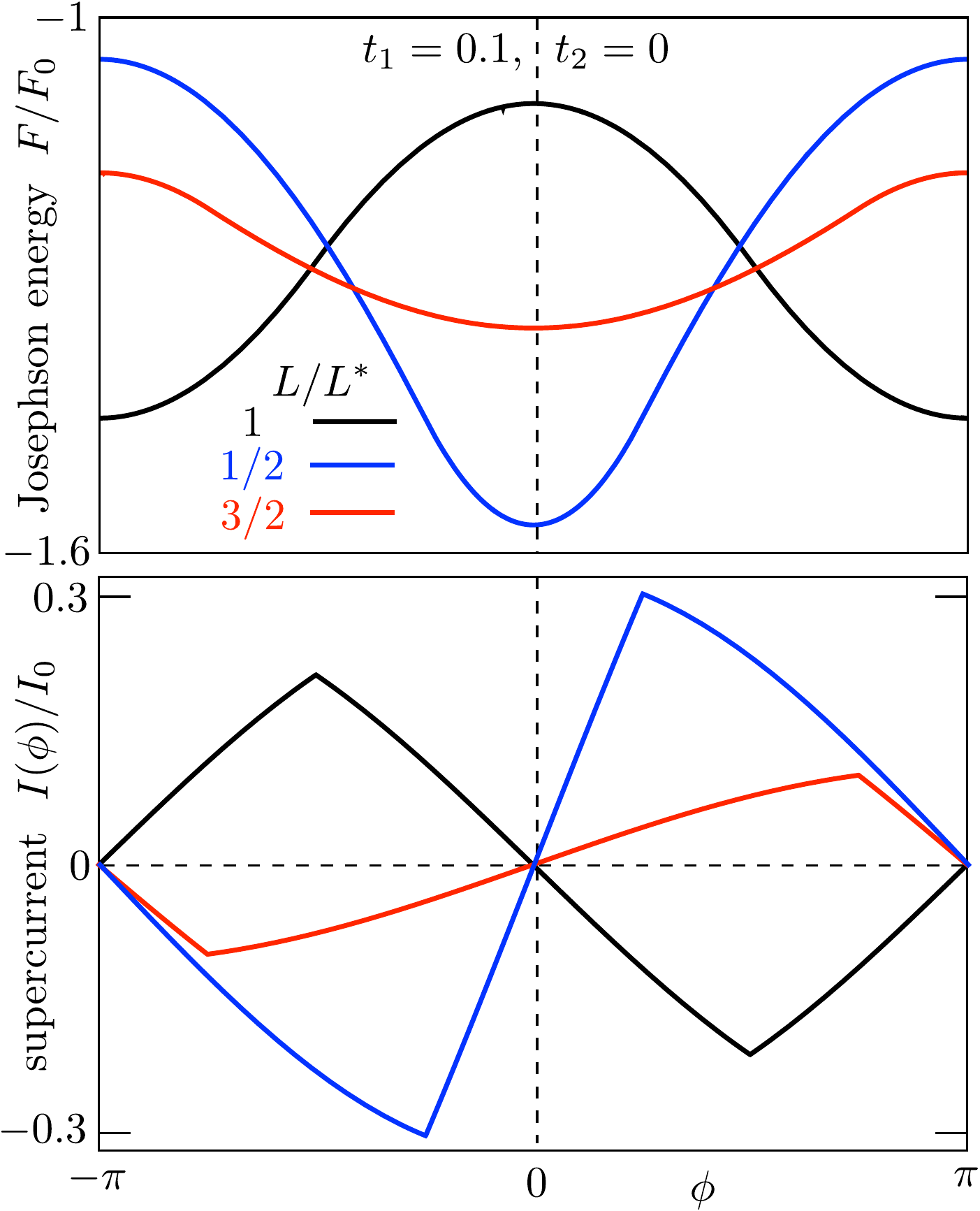}}
\caption{Phase dependence of the Josephson energy (top panel, in units of $F_0=\Delta_0 k_{\rm max}W/\pi$) and the supercurrent (lower panel, in units of $I_0=(e/\hbar)F_0$) in the altermagnet Josephson junction, for different lengths $L$ of the junction [in units of $L^\ast=3\pi/(4t_1 k_{\text{max}})$]. In the interval $2/3<L/L^\ast<4/3$ the Josephson energy is maximal rather than minimal at $\phi=0$, resulting in a negative critical current. These are results for pure $d_{xy}$-wave magnetization ($t_2=0$) and without normal reflection ($\Gamma=1$).
}
\label{fig_iphi1}
\end{figure}

The critical current from Eq.\ \eqref{Inoreflection} is given by
\begin{equation}
I_c=I_0\frac{\sin 2\alpha}{2\alpha},\label{IcGamma1}
\end{equation}
see Fig.\ \ref{fig_iphi1}. A negative sign of $I_c$ means that the maximum supercurrent is reached in the interval $-\pi<\phi<0$. When $I_c<0$ the Josephson energy is minimal at $\phi=\pi$ rather than at $\phi=0$, the altermagnet Josephson junction has become a $\pi$-junction \cite{Oua23,Zha23}. 

With increasing $L$ a negative critical current first appears in the interval $\pi/2<{t_1} k_{\text{max}}L<\pi$. At $L=L^\ast\equiv 3\pi/(4{t_1} k_{\text{max}})$ one has $I_c=-(2/3\pi)I_0$, so the negative $I_c$ is comparable in magnitude to the value $I_0$ at $L=0$. Note that this characteristic length $L^\ast$ is still in the short-junction regime provided that ${t_1}$ is not too small, we need $\Delta_0/\mu\ll{t_1}\ll 1$.

\begin{figure}[tb]
\centerline{\includegraphics[width=0.8\linewidth]{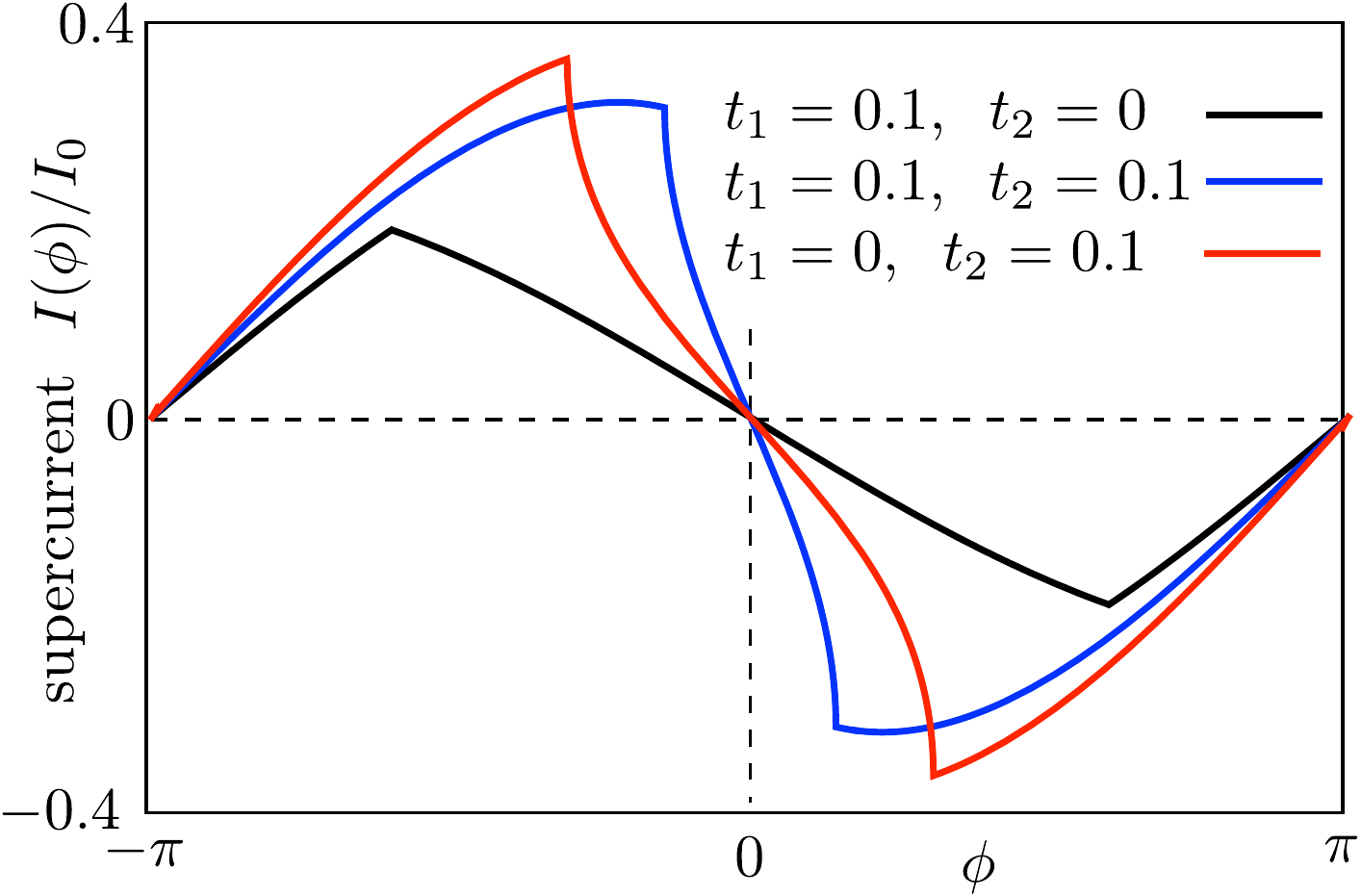}}
\caption{Current-phase relationship for the case of pure $d_{xy}$-wave magnetization (black, $t_2=0$, $t_1=0.1$, $k_{\rm F}L=25$), pure $d_{x^2-y^2}$-wave magnetization (red, $t_1=0$, $t_2=0.1$, $k_{\rm F}L=20$) and the equal weight case (blue, $t_1=t_2=0.1$, $k_{\rm F}L=15$).
}
\label{fig_iphi1b}
\end{figure}

These results are not changed qualitatively if we include a $d_{x^2-y^2}$ contribution to the magnetization, see Fig.\ \ref{fig_iphi1b}. In the case $t_1=0$ of a pure $d_{x^2-y^2}$-wave magnetization the negative critical current first appears in the interval $\pi/2<\tilde{t}_2 k_{\text{max}}L<\pi$, with 
\begin{equation}
\tilde{t}_2=\tfrac{1}{2}(1-2t_2)^{-1/2}-\tfrac{1}{2}(1+2t_2)^{-1/2}=t_2+{\cal O}(t_2^2).
\end{equation}

\begin{figure}[tb]
\centerline{\includegraphics[width=0.8\linewidth]{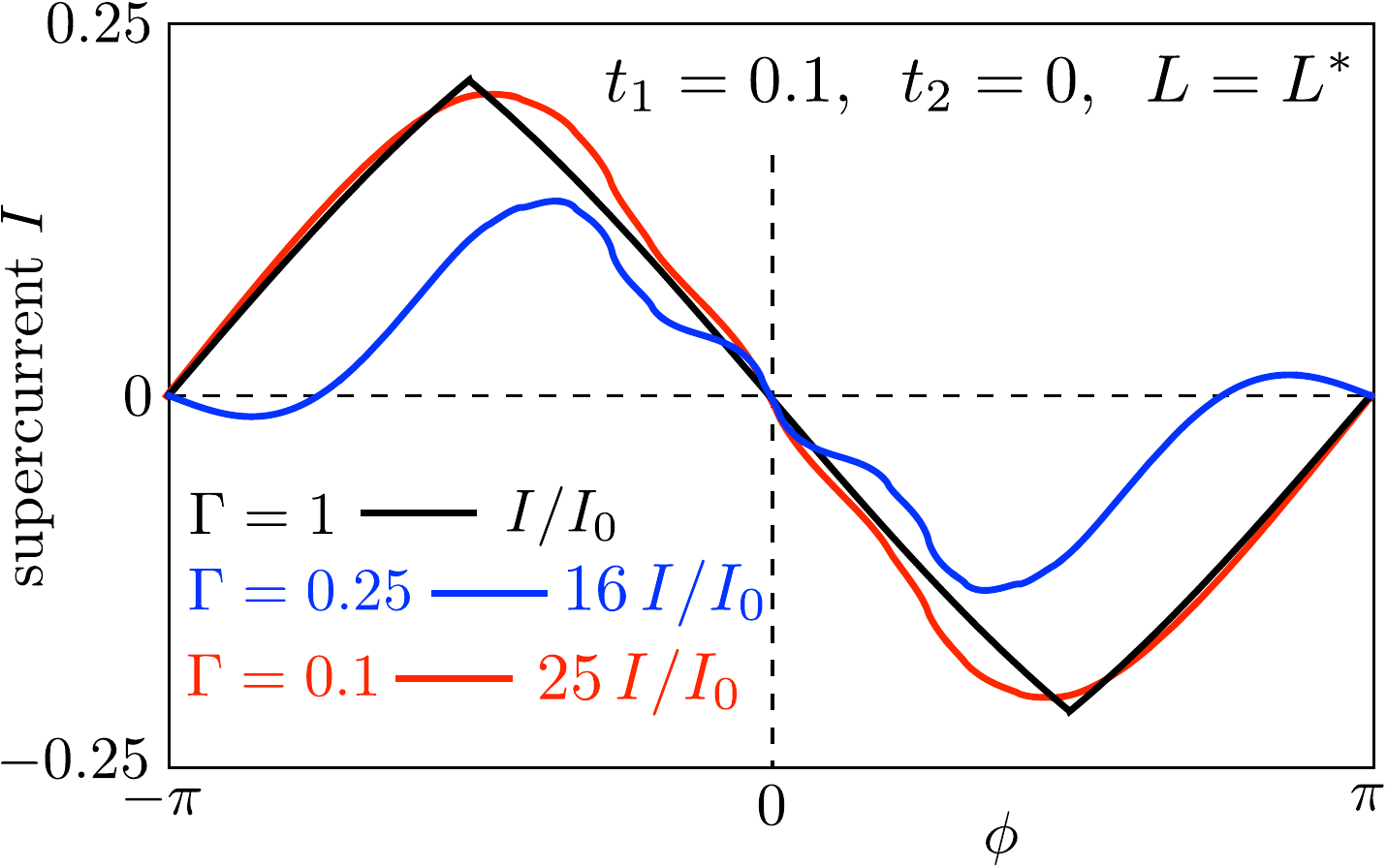}}
\caption{The black curve shows the supercurrent \eqref{Inoreflection} without normal reflections, the blue and red curves include normal reflections from a barrier at each AS interface [transmission probability $\Gamma$, Andreev levels given by Eqs.\ \eqref{Ebarrier}) and \eqref{Tbarrier}]. Each curve is for the same junction length $L=L^\ast$ and $d_{xy}$ magnetization strength $t_1=0.1$ (with $t_2=0$). 
}
\label{fig_iphi2}
\end{figure}

All of this was without normal reflections. We consider the effect of a tunnel barrier (transmission probability $\Gamma$) at each AS interface in the case of pure $d_{xy}$-wave magnetization, when we have the closed-form expression \eqref{Ebarrier} for the Andreev levels. As illustrated in Fig.\ \ref{fig_iphi2}, the barrier reduces the magnitude of the critical current, but its sign remains unchanged. For $\Gamma\ll 1$ the critical current \eqref{IcGamma1} is reduced by a factor $\Gamma$, because only the transmission resonance peaks (unit height, width $\Gamma$) contribute.

We can compare our result \eqref{Inoreflection} for the supercurrent-phase relationship with Ref.\ \cite{Zha23}, which studies the same system in a different formulation. While qualitatively we find the same sign changes in the critical current with increasing $L$, the decay rate of the oscillations is different. Moreover, Eq.\ \eqref{Inoreflection} is strongly nonsinusoidal, while Ref.\ \cite{Zha23} finds $I(\phi)\propto\sin\phi$. The absence of higher harmonics suggests a perturbative approximation. We emphasise that our result is fully non-perturbative. 

\section{Conclusion}

In summary, we have extended the scattering theory of nonmagnetic Josephson junctions to the case of an altermagnetic junction. The basic effect of the \textit{d}-wave magnetization is to spin-polarize the Andreev levels, by giving the spin-up and spin-down spectra $E(\phi)$ opposite phase shifts $\pm\delta\phi$. 

For a planar junction aligned along the crystalline axes of $d_{xy}$-wave symmetry the phase shift is proportional to the transverse momentum $k_y$, which upon integration of $dE/d\phi$ over $k_y$ gives the simple closed-form result \eqref{Inoreflection} for the supercurrent.

As noticed previously \cite{Oua23, Zha23}, the altermagnet Josephson junction undergoes $0$--$\pi$ transitions with increasing junction length $L$, where the critical current changes sign. Our approach is non-perturbative in the transmission probability through the junction, producing the strongly non-sinusoidal current-phase relationship of Figs.\ \ref{fig_iphi1} and \ref{fig_iphi1b}. To observe this the junction length $L$ should be below the superconducting coherence length $\xi_0$ and also below the mean free path $l$ for impurity scattering. The characteristic value of $L$ for a negative critical current is $3\pi/(4t_1k_{\rm F})$, which for $t_1\simeq 0.1$ and $k_{\rm F}\simeq 10^9\,{\rm m}^{-1}$ amounts to a realistically short junction length of $L\simeq 20\,{\rm nm}$.

\acknowledgments

This project has received funding from the European Research Council (ERC) under the European Union's Horizon 2020 research and innovation programme. Advice from A. R. Akhmerov, B. van Heck, and G. Lemut has been very helpful.

\appendix

\section{Scattering matrix calculation of the Andreev levels}
\label{app_barrier}

We consider the altermagnet Josephson junction of Fig.\ \ref{fig_layout}, with a tunnel barrier (transmission probability $\Gamma$) at $x=0$ and $x=L$. We calculate the Andreev level spectrum by means of the scattering formulation of Ref.\ \onlinecite{Bee91}.

The scattering matrix $S(E)$ of electrons (\textit{e}) and holes (\textit{h}) at energy $E$, incident on the altermagnet from the left (\textit{L}) or the right (\textit{R}) with transverse momentum $k_y$, has the block-diagonal form
\begin{equation}
\begin{split}
&S(E)=\begin{pmatrix}
S_e(E)&0\\
0&S_h(E)
\end{pmatrix},\\
& \Psi_{\text{out}}=S\Psi_{\text{in}},\;\;\Psi=(\psi_{e,L},\psi_{e,R},\psi_{h,L},\psi_{h,R}).
\end{split}
\end{equation}
Without the tunnel barrier and at the Fermi level ($E=0$) one would have simply
\begin{subequations}
\begin{align}
&S_e(0)=\begin{pmatrix}
0&\exp(-iL[Q_0^-+Q_z^-])\\
\exp(iL [Q_0^++Q_z^+])&0
\end{pmatrix},\\
&S_h(0)=\begin{pmatrix}
0&\exp(-iL[Q_0^+-Q_z^+])\\
\exp(iL[Q_0^--Q_z^-])&0
\end{pmatrix},
\end{align}
\end{subequations}
in terms of the momentum offsets defined in Eq.\ \eqref{Q0zpm}.

Multiple reflections by the two barriers change this into
\begin{widetext}
\begin{subequations}
\label{Sehdef}
\begin{align}
&S_e(0)=\frac{1}{1+(1-\Gamma)e^{iLK_e}}\begin{pmatrix}
\sqrt{1-\Gamma}(1+ e^{iLK_e})&\Gamma\exp(-iL[Q_0^-+Q_z^-])\\
\Gamma\exp(iL [Q_0^++Q_z^+])&-\sqrt{1-\Gamma}(1+ e^{iLK_e})
\end{pmatrix},\\
&K_e=Q_0^+-Q_0^-+Q_z^+-Q_z^-,\\
&S_h(0)=\frac{1}{1+(1-\Gamma)e^{iLK_h}}\begin{pmatrix}
\sqrt{1-\Gamma}(1+ e^{iLK_h})&\Gamma\exp(-iL[Q_0^+-Q_z^+])\\
\Gamma\exp(iL[Q_0^--Q_z^-])&-\sqrt{1-\Gamma}(1+ e^{iLK_h})
\end{pmatrix},\\
&K_h=-Q_0^++Q_0^-+Q_z^+-Q_z^-.
\end{align}
\end{subequations}
\end{widetext}

We set $\Delta=\Delta_0 e^{i\phi/2}$ at the left superconductor ($x<0$) and $\Delta=\Delta_0e^{-i\phi/2}$ at the right superconductor ($x>L$). The condition for a bound state is
\begin{equation}
{\rm Det}\,[1-R(E)S(E)]=0,\label{Detdef}
\end{equation}
in terms of the scattering matrix $S(E)$ of the normal region and the Andreev reflection matrix
\begin{equation}
\begin{split}
&R(E)=e^{-i\alpha(E)}\begin{pmatrix}
0&R_{eh}\\
R_{he}&0
\end{pmatrix},\\
&R_{eh}=R_{he}^\ast=\begin{pmatrix}
e^{i\phi/2}&0\\
0&e^{-i\phi/2}
\end{pmatrix}.
\end{split}
\label{REdef}
\end{equation}

The function $\alpha(E)$, given by Eq.\ \eqref{alphadef}, varies on the scale of $\Delta_0$. The energy scale on which $S(E)$ varies is of order $\Gamma\bar{v}/L$. If $L\ll\Gamma\bar{v}/\Delta_0$ it is consistent to evaluate $S(E)$ at $E=0$, while retaining the energy dependence of $R(E)$. 

Substitution of of Eqs.\ \eqref{Sehdef} and \eqref{REdef} into Eq.\ \eqref{Detdef} gives the determinantal equation \eqref{Andreevbarrier} from the main text.

\section{Tight-binding calculations}
\label{app_simulations}

For the computer simulations we discretized the altermagnet Hamiltonian \eqref{H0def} on a square lattice (lattice constant $a$, mass $m$, and $\hbar$ all set to unity),
\begin{align}
&H_0=2-\cos k_x-\cos k_y-\mu\nonumber\\
&\quad+t_{1}\sin k_x\sin k_y\sigma_z+2t_{2}(\cos k_x-\cos k_y)\sigma_z.
\end{align}
In the two superconducting regions we set $t_1=0=t_2$ and couple the electron and hole blocks by a pair potential $\Delta_0e^{\pm i\phi/2}$. We keep the same chemical potential $\mu$ throughout. 

The system is infinitely extended in the $y$-direction. In the $x$-direction the altermagnet is in the interval $0<x<L$, while the superconductors occupy the regions $-L_S<x<0$ and $L<x<L_S$. The length $L_S$ is chosen much larger than the superconducting coherence length $\xi_0=\sqrt{2\mu}/\Delta_0$. We took $\mu=0.5$ and $\Delta_0=5\cdot 10^{-4}$, hence $\xi_0=2000$. With $L= 20$ we are therefore deep in the short-junction regime. The tight-binding model is implemented by means of the Kwant toolbox \cite{kwant}.

For Fig.\ \ref{fig_test} we inserted a tunnel barrier at $x=0$ and $x=L$, by locally modifying the hopping matrix elements. The transmission probability $T$ through the junction was calculated separately for $\Delta_0=0$, so that there are no adjustable parameters in the comparison with the analytics.

\end{document}